# Normal modes of coupled vortex gyration in two spatially separated magnetic nanodisks

Ki-Suk Lee, Hyunsung Jung, Dong-Soo Han, and Sang-Koog Kim [a)]

*National Creative Research Center for Spin Dynamics & Spin-Wave Devices, and Nanospinics Laboratory, Research Institute of Advanced Materials, Department of Materials Science and Engineering, Seoul National University, Seoul 151-744, Republic of Korea*

We found from analytical derivations and micromagnetic numerical simulations that there exist two distinct normal modes in apparently complex vortex gyrotropic motions in two dipolar-coupled magnetic nanodisks. The normal modes have characteristic higher and lower single angular eigenfrequencies with their own elliptical orbits elongated along the *x* (bonding axis) and *y* axes, respectively. The superposition of the two normal modes results in coupled vortex gyrations, which depend on the relative vortex-state configuration in a pair of dipolar-coupled disks. This normal-mode representation is a simple means of understanding the observed complex vortex gyrations in two or more dipolar-interacting disks of various vortex-state configurations.

[a)] To whom all correspondence should be addressed; Electronic mail: sangkoog@snu.ac.kr.

The magnetic vortex, which is formed by out-of-plane vortex-core magnetization together with in-plane curling magnetization,[1] has nontrivial low-frequency translational modes (in-plane orbital motion around its equilibrium position) in micron-size or smaller magnetic dots.[2-8] This unique dynamic characteristic of the magnetic vortex has attracted growing interest in that, owing to persistent vortex-core oscillatory motion (i.e. vortex gyration) it can be implemented in nano-oscillators.[7,8] As extension of intensive studies on isolated single vortex-state dots, there have been studies on dynamics of coupled vortex-state dots because, when spatially separated magnetic dots are sufficiently close to each other, dipolar (magnetostatic) interaction affects the vortex excitations, particularly, gyration of individual disks.[9-15] A common finding in earlier studies on the effect of neighboring disks' dynamic dipolar interaction on vortex gyrations is emergent frequency splitting. Shibata *et al.*[9] have analyzed such frequency splitting in a pair of vortices as well as in a two-dimensional array of same. Recently, several experimental observations[10-15] on the gyrations of dipolar-coupled vortices, for example, resonance-frequency broadening[10] in arrays of disks, and asymmetric resonance-frequency splitting[11] in a pair of vortices, have been reported. Additionally, vortex-core gyrations and their asymmetric eigenfrequency splittings have been examined by the present authors.[14,15] Nonetheless, a comprehensive understanding of the fundamentals of dipolar-coupled gyrations remains elusive.

In this article, we report on analytical derivations of the normal modes and their dependences on the relative vortex-state configuration in both disks. We also provide a simple means of understanding apparently complex coupled vortex gyrations in terms of the superposition of the two normal modes, which were also studied by micromagnetic numerical simulations. In the present study, as part of our investigation of coupled vortex gyrations, we conducted micromagnetic simulations of the magnetization dynamics in two identical Permalloy (Py: $Ni_{81}Fe_{19}$) disks of $2R$ = 303 nm diameter, $L$ = 20 nm thickness, and 15 nm edge-to-edge inter-distance. We utilized the OOMMF code[16] that employs the Landau-Lifshitz-Gilbert (LLG) equation.[17] The Py material parameters were applied as described in Ref.[18]. In the model, four different relative vortex-state configurations were utilized, as shown in Fig. 1(a) and as represented by [$p_1$, $C_1$] along with [$p_2$, $C_2$]=[+1,+1], where $p$ = +1(−1) corresponds to the upward (downward) core orientation, and $C$ = +1(−1), the counter-clockwise (clockwise) in-plane curling magnetization. The number in subscript indicates either disk 1 or disk 2. In order to excite all of the modes existing in the two dipolar-coupled disks, the vortex core only in disk 2 (the right disk of each pair) was intendedly displaced to an initial position, 69 nm in the +$y$ direction by application of a 300 Oe field in the +$x$ direction locally,[19] after which both disks were relaxed.

Figures 1(b)-1(d) show the characteristic dynamics of the coupled vortex gyrations for the indicated representative configurations. In all of the cases, the common features were the beating patterns of the oscillatory $x$ and $y$ components of both vortex-core position vectors along with the crossovers between the local maxima and minima of the modulation envelopes between disk 1 and disk 2. In two of our earlier studies,[14,15] these patterns and crossovers

were observed experimentally in the case of $[p_1, C_1] = [-1, +1]$ and $[p_2, C_2] = [+1, +1]$, for example. The beating frequencies [Fig. 1(b)], relative rotation sense and phase differences [Fig. 1(c)] between disk 1 and disk 2, as well as the frequency splitting [Fig. 1(d)], were in contrast with the vortex-state configuration in disk 1 with respect to that in disk 2, where $[p_2, C_2] = [+1, +1]$ was maintained.

In order to fully understand such apparently complex gyrations as found in those simulation results, we analytically derived the normal modes of different single eigenfrequencies, which modes are nontrivial in the case of coupled vortex oscillators in which two different dipolar-coupled vortex cores gyrate. In the analytical derivations, we started with two coupled linearized Thiele's equations,[20]

$$-\mathbf{G}_1 \times \dot{\mathbf{X}}_1 - \hat{D}\dot{\mathbf{X}}_1 + \partial W(\mathbf{X}_1, \mathbf{X}_2)/\partial \mathbf{X}_1 = 0 \qquad (1a)$$

$$-\mathbf{G}_2 \times \dot{\mathbf{X}}_2 - \hat{D}\dot{\mathbf{X}}_2 + \partial W(\mathbf{X}_1, \mathbf{X}_2)/\partial \mathbf{X}_2 = 0 \qquad (1b)$$

where $\mathbf{X}_i = (x_i, y_i)$ is the vortex-core position vector from the center of disk $i$, and where $i = 1$ and 2, $\mathbf{G}_i = -G p_i \hat{\mathbf{z}}$ is the gyrovector with constant $G = 2\pi L M_s / \gamma > 0$ (the saturation magnetization $M_s$ and the gyromagnetic ratio $\gamma$), and $\hat{D} = D\hat{I}$ is the damping tensor with the identity matrix $\hat{I}$ and the damping constant $D$.[6] The total potential energy is given as $W(\mathbf{X}_1, \mathbf{X}_2) = W(0) + \kappa(\mathbf{X}_1^2 + \mathbf{X}_2^2) + W_{\text{int}}$, where $W(0)$ is the potential energy for $\mathbf{X}_i = (0, 0)$, the second term is that for the shifted cores with the identical stiffness coefficient $\kappa$ for the isolated disks,[2] and $W_{\text{int}}$ is the interaction energy for both disks with displaced cores.

Assuming a rigid vortex model[2] and considering only the side-surface charges of each disk, $W_{int}$ can be written simply as $C_1C_2(\eta_x x_1 x_2 - \eta_y y_1 y_2)$, as reported in Ref. 9, where $\eta_x$ and $\eta_y$ represent the interaction strengths along the $x$ and $y$ axes, respectively, and are functions of the inderdistance.[19]

In order to derive the analytical expression of the normal modes of coupled vortex gyrations in a given system, we employed coordinate transformations based on the in-phase and out-of-phase relations between $\mathbf{X}_1$ and $\mathbf{X}_2$ along the $x$ and $y$ axes, respectively, according to the experimental observation reported in our earlier work[15]: we set the two normal-mode coordinates as $\boldsymbol{\Xi} = (x_1+x_2,\ y_1+p_1 p_2 y_2)$ and $\boldsymbol{\Omega} = (x_1-x_2,\ y_1-p_1 p_2 y_2)$. The product of $p_1$ and $p_2$ determines the phase relation in the $y$ component between the two disks for each mode. Through the diagonalization of Eqs. (1a) and (1b) with respect to the normal-mode coordinates, we can obtain these two uncoupled equations of vortex gyrotropic motion,

$$-\begin{bmatrix} D & p_1|G| \\ -p_1|G| & D \end{bmatrix}\dot{\boldsymbol{\Xi}} + \kappa \begin{bmatrix} 1+\Gamma_x & 0 \\ 0 & 1-\Gamma_y \end{bmatrix}\boldsymbol{\Xi} = 0 \qquad (2a)$$

$$-\begin{bmatrix} D & p_1|G| \\ -p_1|G| & D \end{bmatrix}\dot{\boldsymbol{\Omega}} + \kappa \begin{bmatrix} 1-\Gamma_x & 0 \\ 0 & 1+\Gamma_y \end{bmatrix}\boldsymbol{\Omega} = 0 \qquad (2b),$$

where $\Gamma_x = C_1 C_2 \eta_x / \kappa$ and $\Gamma_y = C_1 C_2 p_1 p_2 \eta_y / \kappa$. The general solutions of Eqs. (2a) and (2b) are written simply as $\boldsymbol{\Xi} = \boldsymbol{\Xi}_0 \exp[-i(\tilde{\omega}_\Xi t + \varphi_\Xi)]$ and $\boldsymbol{\Omega} = \boldsymbol{\Omega}_0 \exp[-i(\tilde{\omega}_\Omega t + \varphi_\Omega)]$ with the corresponding amplitude vectors of $\boldsymbol{\Xi}_0 = (\Xi_{0x}, \Xi_{0y})$ a.nd $\boldsymbol{\Omega}_0 = (\Omega_{0x}, \Omega_{0y})$ and the phase constants of $\varphi_\Xi$ and $\varphi_\Omega$. By inserting these general solutions into Eqs. (2a) and (2b), we can obtain analytical expressions of the complex angular frequencies $\tilde{\omega}_\Xi$ and $\tilde{\omega}_\Omega$ as well as the ratios of $\Xi_{0y}/\Xi_{0x}$ and $\Omega_{0y}/\Omega_{0x}$, for the normal modes. On the basis of the relation between the ordinary and the normal-mode coordinates, that is, $\mathbf{X}_1 = \frac{1}{2}(\Xi_x + \Omega_x, \Xi_y + \Omega_y)$ and $\mathbf{X}_2 = \frac{1}{2}(\Xi_x - \Omega_x, p_1 p_2 \Xi_y - p_1 p_2 \Omega_y)$, the normal modes of coupled vortex-core gyrations, in the ordinary coordinates, can be derived analytically: $\mathbf{X}_{1,\Xi} = \frac{1}{2}\Xi_0 \exp[-i(\tilde{\omega}_\Xi t + \varphi_\Xi)]$, $\mathbf{X}_{2,\Xi} = \frac{1}{2}\Xi'_0 \exp[-i(\tilde{\omega}_\Xi t + \varphi_\Xi)]$ and $\mathbf{X}_{1,\Omega} = \frac{1}{2}\Omega_0 \exp[-i(\tilde{\omega}_\Omega t + \varphi_\Omega)]$, $\mathbf{X}_{2,\Omega} = \frac{1}{2}\Omega'_0 \exp[-i(\tilde{\omega}_\Omega t + \varphi_\Omega)]$ with the corresponding angular eigenfrequencies $\mathrm{Re}(\tilde{\omega}_\Xi)$ and $\mathrm{Re}(\tilde{\omega}_\Omega)$, where $\Xi'_0 = (\Xi_{0x}, p_1 p_2 \Xi_{0y})$ and $\Omega'_0 = -(\Omega_{0x}, p_1 p_2 \Omega_{0y})$. The general solutions of Eqs. (1a) and (1b) can also be given, by the superposition of the two normal modes in disk 1 and disk 2, such that $\mathbf{X}_1 = \mathbf{X}_{1,\Xi} + \mathbf{X}_{1,\Omega}$ and $\mathbf{X}_2 = \mathbf{X}_{2,\Xi} + \mathbf{X}_{2,\Omega}$.

For the cases of $|G| \gg |D|$ and $\eta_x, \eta_y \ll \kappa$, $\tilde{\omega}_\Xi$ and $\tilde{\omega}_\Omega$ approximate to be $\tilde{\omega}_\Xi \approx \omega_0 \left( \sqrt{(1+\Gamma_x)(1-\Gamma_y)} + iD/G \right)$ and $\tilde{\omega}_\Omega \approx \omega_0 \left( \sqrt{(1-\Gamma_x)(1+\Gamma_y)} + iD/G \right)$ with the angular eigenfrequency of vortex gyration in an isolated disk, $\omega_0 = \kappa/|G|$.[2] The angular eigenfrequencies of the uncoupled $\Xi$ and $\Omega$ normal modes are simply rewritten as

$$\mathrm{Re}(\tilde{\omega}_\Xi) \approx \omega_0 \sqrt{(1 + C_1 C_2 \eta_x/\kappa)(1 - C_1 C_2 p_1 p_2 \eta_y/\kappa)}$$

and

$$\mathrm{Re}(\tilde{\omega}_\Omega) \approx \omega_0 \sqrt{(1 - C_1 C_2 \eta_x/\kappa)(1 + C_1 C_2 p_1 p_2 \eta_y/\kappa)},$$

respectively. Consequently, the angular

frequency difference $\Delta\omega$, defined as Re($\tilde{\omega}_\Xi$) − Re($\tilde{\omega}_\Omega$), is expressed as $\omega_0 C_1 C_2(\eta_x - p_1 p_2 \eta_y)/\kappa$, which is determined by the value of $C_1 C_2 p_1 p_2$. For the case where the $x$ axis is the bonding axis, $\eta_y > \eta_x$ always holds, such that $\Delta\omega < 0$ for $C_1 C_2 p_1 p_2 = +1$ and $\Delta\omega > 0$ for $C_1 C_2 p_1 p_2 = -1$. However, the magnitude of the angular frequency splitting $|\Delta\omega|$ is determined by only $p_1 p_2$, and $\left|\Delta\omega_{p_1 p_2 = +1}\right| = \omega_0(\eta_y - \eta_x)/\kappa < \left|\Delta\omega_{p_1 p_2 = -1}\right| = \omega_0(\eta_y + \eta_x)/\kappa$, as confirmed by the simulation results shown in Fig. 1(d), which are consistent with the analytical results reported in Ref. 9.

The shapes of the orbital trajectory of the $\Xi$ and $\Omega$ modes can be estimated as $\left|\Xi_{0y}/\Xi_{0x}\right| = \sqrt{(\kappa + C_1 C_2 \eta_x)/(\kappa - C_1 C_2 p_1 p_2 \eta_y)}$ and $\left|\Omega_{0y}/\Omega_{0x}\right| = \sqrt{(\kappa - C_1 C_2 \eta_x)/(\kappa + C_1 C_2 p_1 p_2 \eta_y)}$. Thus, the elongation axis and the degree (hereafter, "ellipticity") of elongations of the normal modes' orbits vary according to the combinations of $C_1 C_2 = \pm 1$ and $p_1 p_2 = \pm 1$ displayed in Table 1. The elongation axis of the normal mode that has the higher angular eigenfrequency (shaded area) is always, for all cases, the bonding ($x$) axis.

To numerically calculate the $\Xi$ and $\Omega$ normal modes using the above analytical expressions, it is necessary to know $\eta_x$ and $\eta_y$ for a given model system. These interaction strengths can be simply estimated through the relations of $\eta_x/\kappa = (|\Delta\omega_-|+|\Delta\omega_+|)/2\omega_0$ and $\eta_y/\kappa = (|\Delta\omega_-|-|\Delta\omega_+|)/2\omega_0$ for the case of $C_1 C_2 = +1$, where $|\Delta\omega_+|$ and $|\Delta\omega_-|$ correspond to the angular frequency splitting for the cases of $p_1 p_2 = +1$ and −1, respectively. For isolated disks of the same geometry and material parameters, we can extract the numerical values of $\eta_x = 1.1\times10^{-4}$ and $\eta_y = 3.1\times10^{-4}$ J/m² from $\kappa = 3\times10^{-3}$ J/m² , $\omega_0 = 2\pi \times 575$ MHz, $|\Delta\omega_+| = 2\pi \times 105$ MHz, and $|\Delta\omega_-| = 2\pi \times 50$ MHz for the case of $C_1C_2 = +1$, as obtained from the simulation results shown in Fig. 1(d).

Using these above values, we calculated $\Xi = \Xi_0 \exp[-i(\tilde{\omega}_\Xi t + \varphi_\Xi)]$ and $\Omega = \Omega_0 \exp[-i(\tilde{\omega}_\Omega t + \varphi_\Omega)]$, where the initial core displacements were set to $(x_2, y_2) = (0, 69$ nm$)$ and $(x_1, y_1) = (0, C_1\times12$ nm$)$ in order to obtain the same initial phases of the core positions as those used in the micromagnetic simulations.[19] The $x$- and $y$-component oscillations of the normal modes, that is, $\Xi_{x,y}$ and $\Omega_{x,y}$, their trajectories, $(\Xi_x, \Xi_y)$ and $(\Omega_x, \Omega_y)$ on the normal-mode coordinates, and their frequency spectra, are plotted in Figs. 2(a), 2(b), and 2(c), respectively, for all of the different combinations of $p_1p_2 = \pm 1$ and $C_1C_2 = \pm1$. The analytical calculations (solid lines) are in quantitative agreement with those (open symbols) extracted from the simulation results plotted in Fig. 1. These comparisons prove that complex coupled vortex-core gyrations in two coupled oscillators such as those shown in Fig. 1 can be predicated or interpreted simply in terms of the superposition of the $\Xi$ and $\Omega$ modes.

Furthermore, the individual contributions of the $\Xi$ and $\Omega$ modes to the gyration of each of disk 1 and 2 can be decomposed into $\mathbf{X}_{1,\Xi}$, $\mathbf{X}_{1,\Omega}$ and $\mathbf{X}_{2,\Xi}$, $\mathbf{X}_{2,\Omega}$, as shown in Fig. 3. The analytical calculations (solid lines) of $\mathbf{X}_{1,\Xi}$, $\mathbf{X}_{1,\Omega}$ and $\mathbf{X}_{2,\Xi}$, $\mathbf{X}_{2,\Omega}$ were in excellent agreements with those (symbols) obtained from the simulation results[21] through the normal-to-ordinary coordinate transformation, as described earlier. Since the superposition of the two normal modes gives rise to the net coupled gyration of each disk (i.e., $\mathbf{X}_1 = \mathbf{X}_{1,\Xi} + \mathbf{X}_{1,\Omega}$ and $\mathbf{X}_2 = \mathbf{X}_{2,\Xi} + \mathbf{X}_{2,\Omega}$), the contrasting eigenfrequencies and phases between the $\Xi$ and $\Omega$ modes, which vary with both $p_1 p_2$ and $C_1 C_2$, determine the modulation frequency (see Fig. 1(b)) and the relative phase of the vortex-core orbital trajectory (see Fig. 1(c)).

The physical origin of the above-noted frequency splitting and complex coupled vortex-core gyrations can be ascribed to the breaking of the radial symmetry of the potential wells of decoupled disks, which is caused by dynamically variable dipolar interaction between those disks, and which depends on the disk pair's relative vortex-state configuration, as explained above.

In summary, we analytically derived two normal modes of coupled vortex-core gyrations in two spatially separated magnetic disks. Dipolar interaction between two such disks breaks the radial symmetry of their potential energy, giving rise to two distinct normal modes, each with a characteristic single eigenfrequency and an elliptical orbit. The frequency splitting and the orbital shape vary with the relative vortex-state configuration. This work provides a simple but complete means of understanding complex vortex gyrations in dipolar-coupled vortex oscillators.

This work was supported by the Basic Science Research Program through the National Research Foundation of Korea (NRF) funded by the Ministry of Education, Science and Technology (grant no. 20100000706).

# Figure captions

**FIG. 1.** (color online) Representative coupled vortex gyrations for the indicated four different polarization ($p$) and chirality ($C$) configurations in a pair of vortex-state disks shown in (a). In (a), the streamlines with the small arrows indicate the in-plane curling magnetizations, and the height displays the out-of-plane magnetizations. (b) The $x$ and $y$ components of the vortex-core position vectors in both disks, as functions of time. (c) Orbital trajectories of the vortex-core gyrations during the time period $t = 0 - 5$ ns. The open circles represent the initial core positions. (d) Frequency spectra obtained from the data shown in (b).

**FIG. 2.** (a) Oscillatory $x$ and $y$ components of the $\Xi$ and $\Omega$ modes, (b) their orbital trajectories, and (c) frequency spectra obtained from the oscillation of the $x$ components of the $\Xi$ and $\Omega$ modes for the four different configurations of $[p_1,C_1]$ with respect to $[p_2,C_2]=[+1, +1]$. The solid lines and open circles correspond to the analytical calculations using Eqs. 2(a) and 2(b) (see text) and the micromagnetic simulation results, respectively.

**FIG. 3.** Contributions of the $\Xi$ and $\Omega$ modes to each disk's vortex-core gyration, i.e., $\mathbf{X}_{1,\Xi}$, $\mathbf{X}_{2,\Xi}$, $\mathbf{X}_{1,\Omega}$, $\mathbf{X}_{2,\Omega}$ (Ref. 21), for the four different configurations of $[p_1,C_1]$ with respect to $[p_2,C_2]=[+1, +1]$. The solid lines and open circles correspond to the analytical calculations (see text) and the micromagnetic simulation results, respectively.

**Table 1**. The major (elongation) axis and the ellipticity (the ratio of the length of the major to that of the minor axis) of each mode for all combinations of $C_1C_2 = \pm 1$ and $p_1p_2 = \pm 1$. The shaded area corresponds to the higher-frequency mode for the given $C_1C_2$ and $p_1p_2$. The numbers in parentheses indicate the numerical values of the ellipticity.

| | | $p_1p_2$ | | | | | | | |
|---|---|---|---|---|---|---|---|---|---|
| | | +1 | | | | -1 | | | |
| | | $\Xi$ mode | | $\Omega$ mode | | $\Xi$ mode | | $\Omega$ mode | |
| | | Major axis | Ellipticity | Major axis | Ellipticity | Major axis | Ellipticity | Major axis | Ellipticity |
| $C_1C_2$ | +1 | $y$ | $\sqrt{\dfrac{\kappa-\eta_y}{\kappa+\eta_x}}$ (0.930) | $x$ | $\sqrt{\dfrac{\kappa-\eta_x}{\kappa+\eta_y}}$ (0.934) | $x$ | $\sqrt{\dfrac{\kappa+\eta_x}{\kappa+\eta_y}}$ (0.969) | $y$ | $\sqrt{\dfrac{\kappa-\eta_y}{\kappa-\eta_x}}$ (0.965) |
| | -1 | $x$ | $\sqrt{\dfrac{\kappa-\eta_x}{\kappa+\eta_y}}$ (0.934) | $y$ | $\sqrt{\dfrac{\kappa-\eta_y}{\kappa+\eta_x}}$ (0.930) | $y$ | $\sqrt{\dfrac{\kappa-\eta_y}{\kappa-\eta_x}}$ (0.965) | $x$ | $\sqrt{\dfrac{\kappa+\eta_x}{\kappa+\eta_y}}$ (0.969) |

Figures

Figure 1.

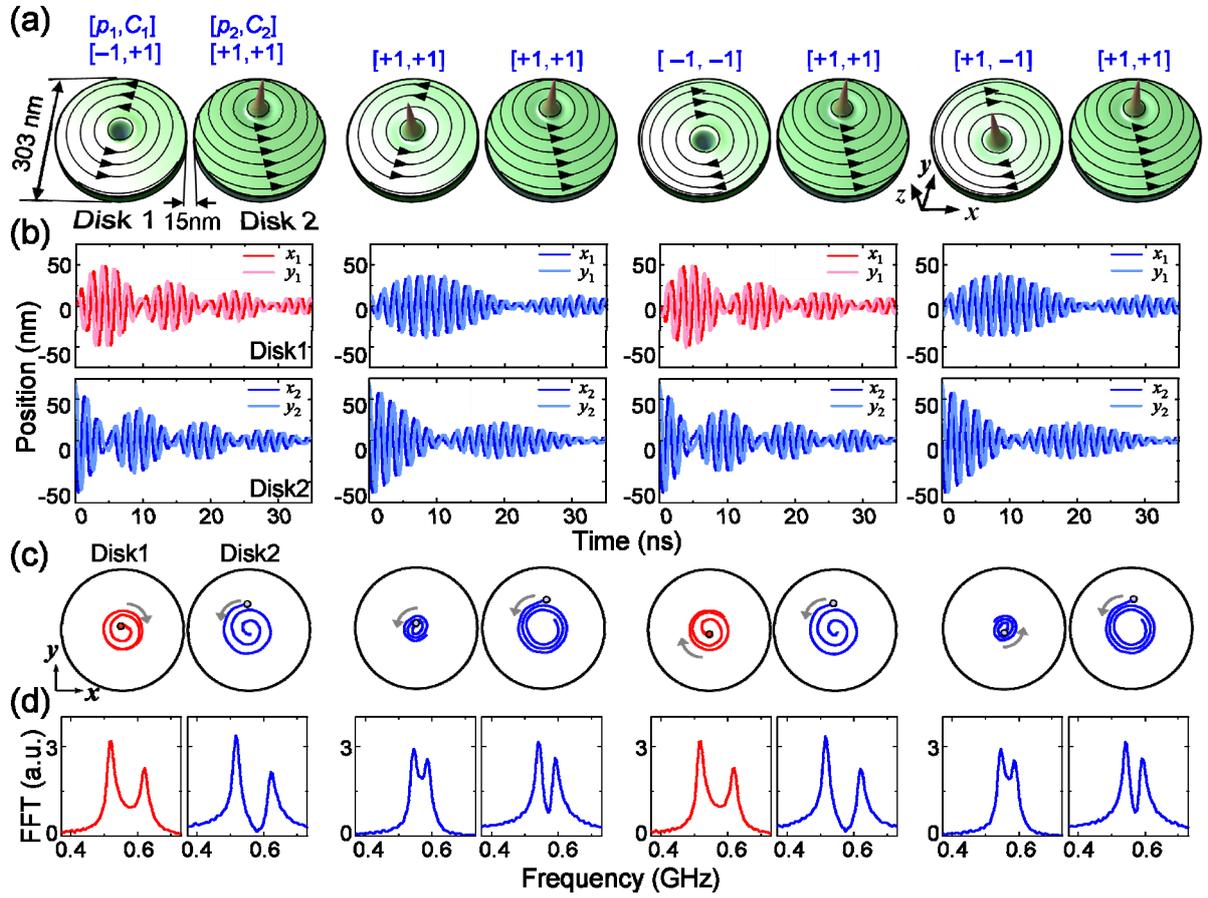

Figure 2.

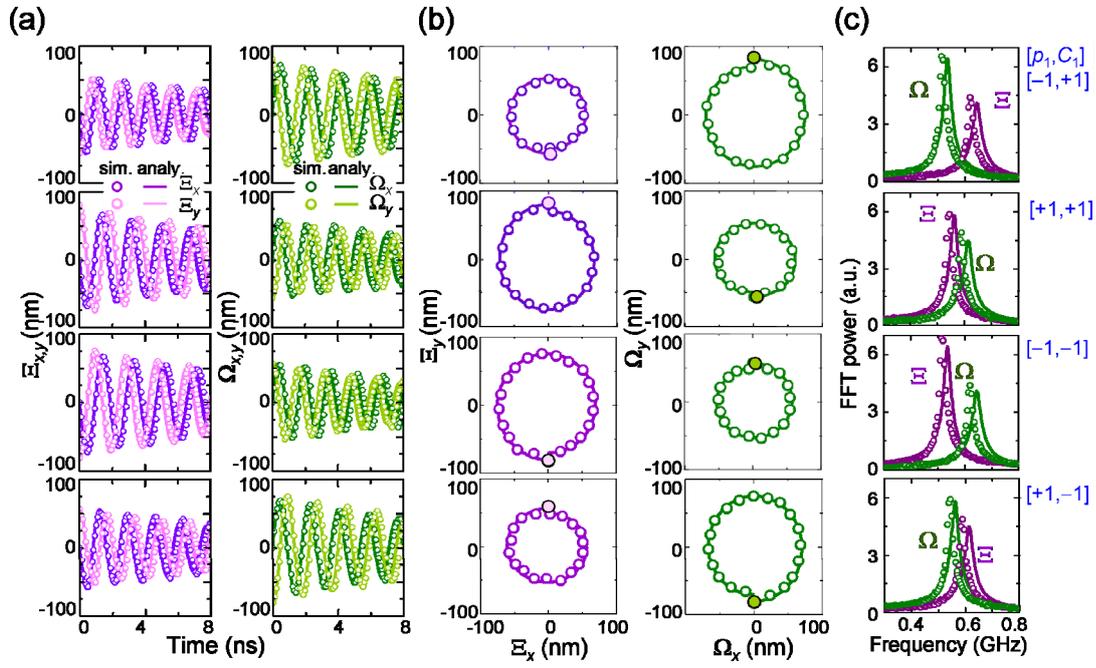

Figure 3.

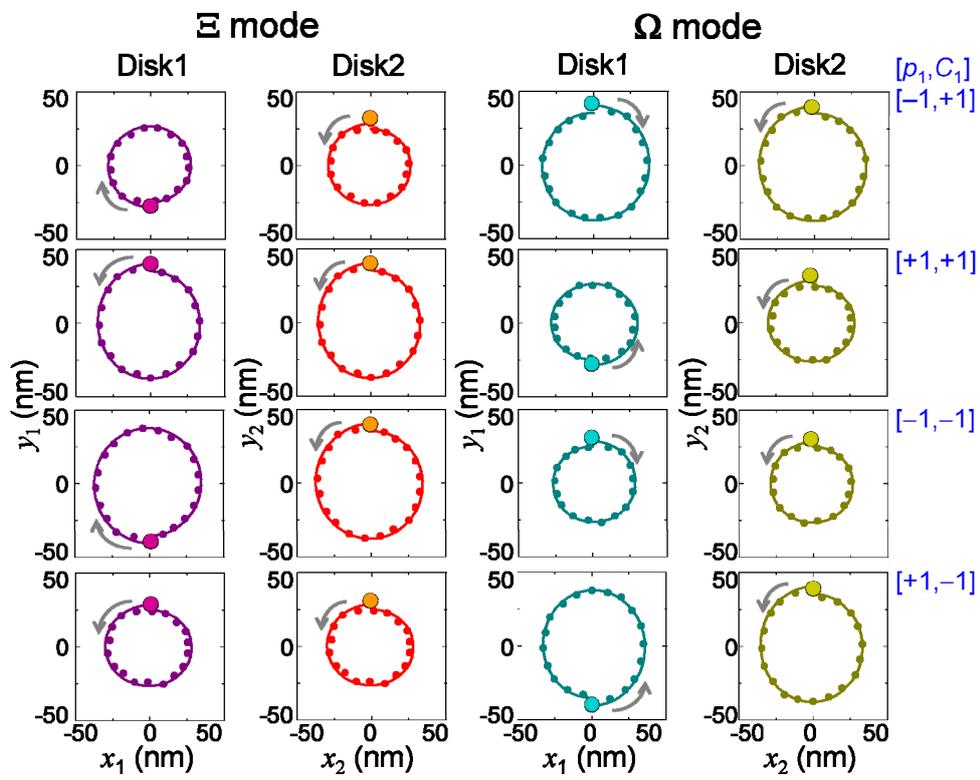